\newif\ifsingle

\newif\ifFullVersion

\documentclass[conference]{IEEEtran}

\usepackage{times}
\usepackage{amsmath,dsfont}
\usepackage{amssymb,amsthm}
\usepackage{epsfig,verbatim}
\usepackage{subfigure}
\usepackage{setspace}
\usepackage{color}
\usepackage{cite}
\usepackage{epstopdf}
\usepackage{graphics}
\usepackage{accents}
\usepackage{acronym}
\usepackage{booktabs}
\usepackage{mathtools}
\usepackage{algorithm}
\usepackage{algorithmic}
\usepackage{enumitem}

\newcommand{\myVec}[1]{{\boldsymbol{#1}}}
\newcommand{\myMat}[1]{{\boldsymbol{#1}}}

\newcommand{\E}{\mathds{E}}		 			
\newcommand{\SigR}{\sigma_r^2}
\newcommand{\SigB}{\sigma_b^2}

\newtheorem{theorem}{Theorem}

\newtheorem{proposition}{Proposition}

\ifsingle

\else

\fi 

\acrodef{adc}[ADC]{Analog-to-Digital Convertor}
\acrodef{dac}[DAC]{digital-to-analog convertor}
\acrodef{cs}[CS]{Compressed Sensing}
\acrodef{dtft}[DTFT]{discrete-time Fourier transform}
\acrodef{dnn}[DNN]{deep neural network} 
\acrodef{csi}[CSI]{channel state information}
\acrodef{map}[MAP]{maximum a-posteriori probability}
\acrodef{snr}[SNR]{Signal-to-Noise Ratio}
\acrodef{sinr}[SINR]{signal-to-interference-and-noise ratio}
\acrodef{bs}[BS]{Base Station} 
\acrodef{iot}[IOT]{Internet of Things}
\acrodef{mimo}[MIMO]{Multiple-Input Multiple-Output}
\acrodef{mse}[MSE]{Mean-Squared Error}
\acrodef{pdf}[PDF]{probability density function}
\acrodef{rv}[RV]{random variable}
\acrodef{tdd}[TDD]{time division duplexing}
\acrodef{rs}[RS]{Reed-Solomon}
\acrodef{lti}[LTI]{linear time-invariant}
\acrodef{wss}[WSS]{wide-sense stationary}
\acrodef{psd}[PSD]{power spectral density}
\acrodef{ser}[SER]{symbol error rate} 
\acrodef{ber}[BER]{bit error rate} 
\acrodef{isi}[ISI]{intersymbol interference}  
\acrodef{awgn}[AWGN]{additive white Gaussian noise} 
\acrodef{ut}[UT]{User Terminal} 
\acrodef{mmw}[mmWave]{millimeter wave}
\acrodef{ris}[RIS]{Reconfigurable Intelligent Surface} 
\acrodef{dma}[DMA]{Dynamic Metasurface Antenna} 

\acrodef{hris}[HRIS]{Hybrid Reconfigurable Intelligent Surface} 

\IEEEoverridecommandlockouts

\newcommand{\vecc}{{\operatorname{vec}}}

\title{Channel Estimation with Simultaneous Reflecting and Sensing Reconfigurable Intelligent Metasurfaces}
\author{Haiyang Zhang$^1$, Nir Shlezinger$^2$,  Idban Alamzadeh$^3$,\\ George C. Alexandropoulos$^4$, Mohammadreza F. Imani$^3$, and Yonina C. Eldar$^1$\\
$^1$Faculty of Math and Computer Science, Weizmann Institute of Science, Rehovot, Israel \\
$^2$School of ECE, Ben-Gurion University of the Negev, Beer-Sheva, Israel\\
$^3$School of ECEE, Arizona State University, Tempe, AZ, USA\\
$^4$Department of Informatics and Telecommunications, National and Kapodistrian University of Athens, Greece\\
\thanks{This work has been supported by the EU H2020 projects BNYQ under grant No. 646804 and RISE-6G under grant No. 101017011, as well as the Israel Science Foundation under grant No. 0100101.
}
\vspace{-0.4cm}
}

\begin{document}
	
	\maketitle
	\pagestyle{empty}
	\thispagestyle{empty}
\begin{abstract}
\acp{ris} are envisioned to play a key role in future wireless communications, enabling programmable radio propagation environments. They are usually considered as nearly passive planar structures that operate as adjustable reflectors, giving rise to a multitude of implementation challenges, including an inherent difficulty in estimating the underlying wireless channels. In this paper, we propose the concept of Hybrid RISs (HRISs), which do not solely reflect the impinging waveform in a controllable fashion, but are also capable of sensing and processing a portion of it via some active reception elements. We first present implementation details for this novel metasurface architecture and propose a simple model for its operation, when considered for wireless communications. As an indicative application of HRISs, we formulate and solve the individual channels identification problem for the uplink of multi-user HRIS-empowered systems. Our numerical results showcase that, in the high signal-to-noise regime, HRISs enable individual channel estimation with notably reduced amounts of pilots, compared to those needed when using a purely reflective \ac{ris} that can only estimate the cascaded channel.


\end{abstract}



\acresetall

\vspace{-0.1cm}
\section{Introduction}
An emerging technology for the future sixth Generation (6G) of wireless communications, enabling dynamically programmable signal propagation over the wireless medium \cite{rise6g}, is the \acp{ris} \cite{huang2019reconfigurable,wu2019towards}. Those planar structures typically consist of multiple metamaterial elements, whose ElectroMagnetic (EM) properties can be externally controlled in a nearly passive manner, allowing them to realize various reflection and scattering profiles \cite{huang2020holographic}.   

The passive nature of RISs implies that they can only act as adjustable reflectors, thus, they can neither receive nor transmit their own data. While \acp{ris} enable smart programmable environments \cite{rise6g}, their purely passive operation also induces notable challenges on the communicating entities. For instance, the introduction of an \ac{ris} implies that a signal transmitted from each \ac{ut} to the \ac{bs} undergoes at least two channels: the \ac{ut}-\ac{ris} and \ac{ris}-\ac{bs} channels. Estimating these individual channels is a challenging task due to the passive nature of \acp{ris}. Consequently, the common approach in the field of RIS-empowered networks is to estimate only the entangled combined effect of these channels, known as the cascaded channel \cite{wang2020channel, liu2019matrix}, which limits the transmission scheme design and restricts network management flexibility \cite{hu2019two}. For example, the individual channels between UT-RIS and RIS-BS are needed for some precoding designs, as discussed in \cite{ye2020joint}. In fact, it was recently proposed to equip \acp{ris} with minimal receive Radio-Frequency (RF) chains to overcome the communication challenges associated with their purely passive counterparts \cite{taha2021enabling, alexandropoulos2020hardware}.

Active metasurfaces have recently emerged as a appealing technology for realizing low-cost and low-power large-scale \ac{mimo} antennas \cite{shlezinger2020dynamic}. \acp{dma} pack large numbers of tunably radiative metamaterials on top of waveguides, resulting in \ac{mimo} transceivers with advanced analog processing capabilities \cite{shlezinger2019dynamic,wang2020dynamic,zhang2021beam}. While the implementation of \acp{dma} differs from passive \acp{ris}, the similarity in the structure of the metamaterial elements between them indicates the feasibility of designing hybrid reflecting and sensing elements. This motivates studying the benefits from such a hybrid metasurface architecture, as an efficient means of facilitating \ac{ris}-empowered wireless communications. 

In this work, we present an initial study on the potential gains of using Hybrid reflecting and sensing \acp{ris} (HRISs) in multi-user wireless communications. To that aim, we begin by discussing the feasibility of the concept of hybrid metamaterials, providing a high-level description of their design. Then, we propose a model for HRISs which captures their ability to simultaneously reflect and receive the incoming signal in an element-by-element controllable manner. To quantify the benefits of HRISs, we study the individual channels estimation problem. Our results show that, in the high \ac{snr} regime, HRISs yield achievable \ac{mse} performance using smaller numbers of pilot signals than those typically required in networks with reflective RISs to estimate solely the cascaded channel \cite{wang2020channel}. Our numerical evaluations also characterize the inherent tradeoff between the ability to estimate the individual channels and tunable reflection, which is balanced by the HRIS configuration.
 
Throughout the paper, we use boldface lower-case and upper-case letters for vectors and matrices, respectively. Calligraphic letters are used for sets. The   vectorization operator,  transpose, conjugation, Hermitian transpose, trace, and  expectation are written as  ${\rm vec}(\cdot)$, $(\cdot)^T$, $(\cdot)^{\dag}$,  $(\cdot)^H$, ${\rm {Tr}}\left(\cdot\right)$, and $\E\{ \cdot \}$,  respectively.
${\rm blkdiag}\left\{{\bf A}_1,{\bf A}_2,\ldots,{\bf A}_n\right\}$ denotes a block diagonal matrix with diagonal blocks given by ${\bf A}_1,{\bf A}_2,\ldots,{\bf A}_n$. 

\section{HRISs and System Modeling}
\label{sec:Model}
\begin{figure}
    \centering
    \includegraphics[width=0.5\columnwidth]{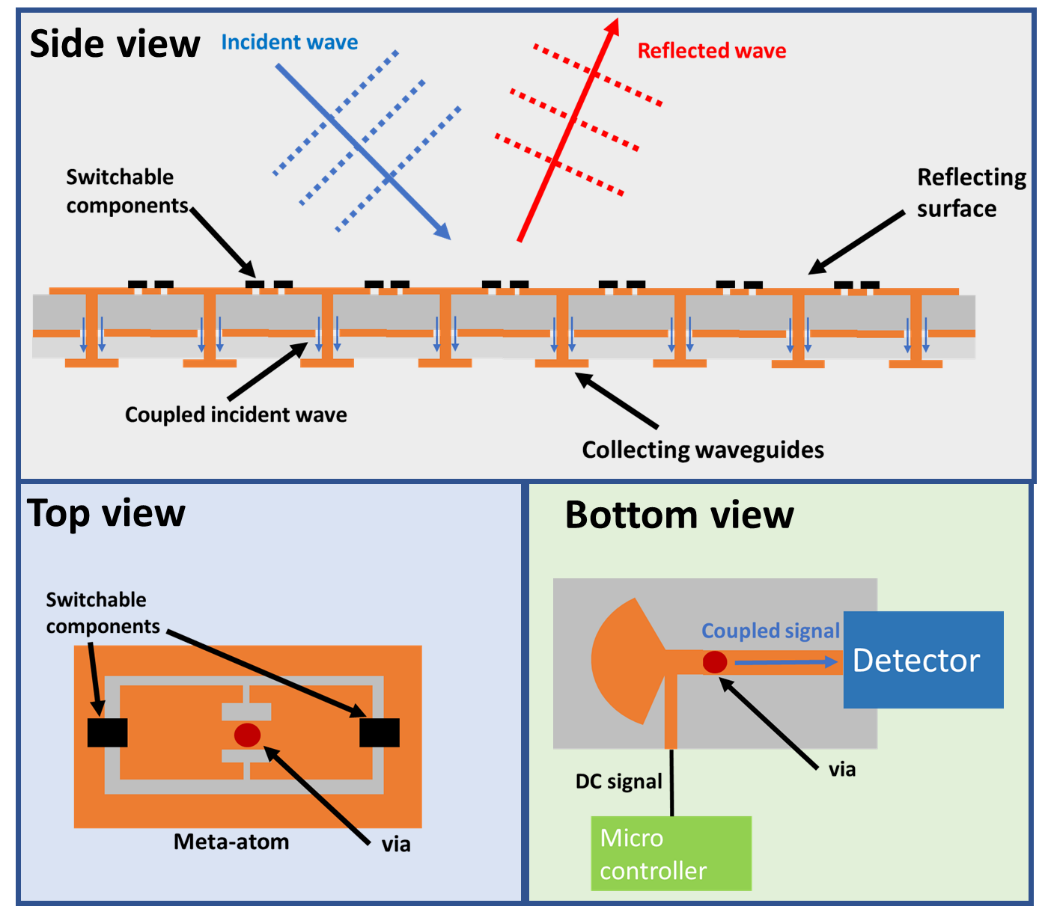}
    \caption{Illustration of the proposed hybrid metamaterials.}
    \label{fig:HybridAtom_v01}
\end{figure}

\subsection{Hybrid Metasurfaces}
\label{subsec:HybridElements}
A rich body of literature has examined the implementation of solely reflective \acp{ris}. A variety of implementations has been recently presented in \cite{huang2020holographic}, ranging from RISs that change the wave propagation inside a multi-scattering environment for improving the received signal, to those which realize anomalous reflection, such that the reflected beam does not follow Snell's law and is directed towards desired directions. In all these efforts, the RIS is not designed to sense the impinging signal. Nonetheless, metasurfaces can be designed to operate in a hybrid reflecting  and sensing manner. Such hybrid operation requires that each metarsurface element is capable of simultaneously reflecting a portion of the impinging signal and receiving a portion of it in a controlable manner. As illustrated in Fig.~\ref{fig:HybridAtom_v01}, a simple mechanism for implementing such an operation is to couple each element to a waveguide. The signals coupled to the waveguides are then measured by receive RF chains and used to determine the necessary information about the channel. A detailed description of the practical implementation of such hybrid metamaterials can be found in \cite{alexandropoulos2021hybrid}. In this paper, we are interested in examining the potential benefits of HRISs in wireless communications. To this end, we present in the sequel a simple model capturing their simultaneous reflecting and sensing operation. 

\vspace{-0.2cm}
\subsection{HRIS Operation Modeling}
\label{subsec:HybridRIS}
\begin{figure}
		\centering			
		\includegraphics[width=0.7\columnwidth]{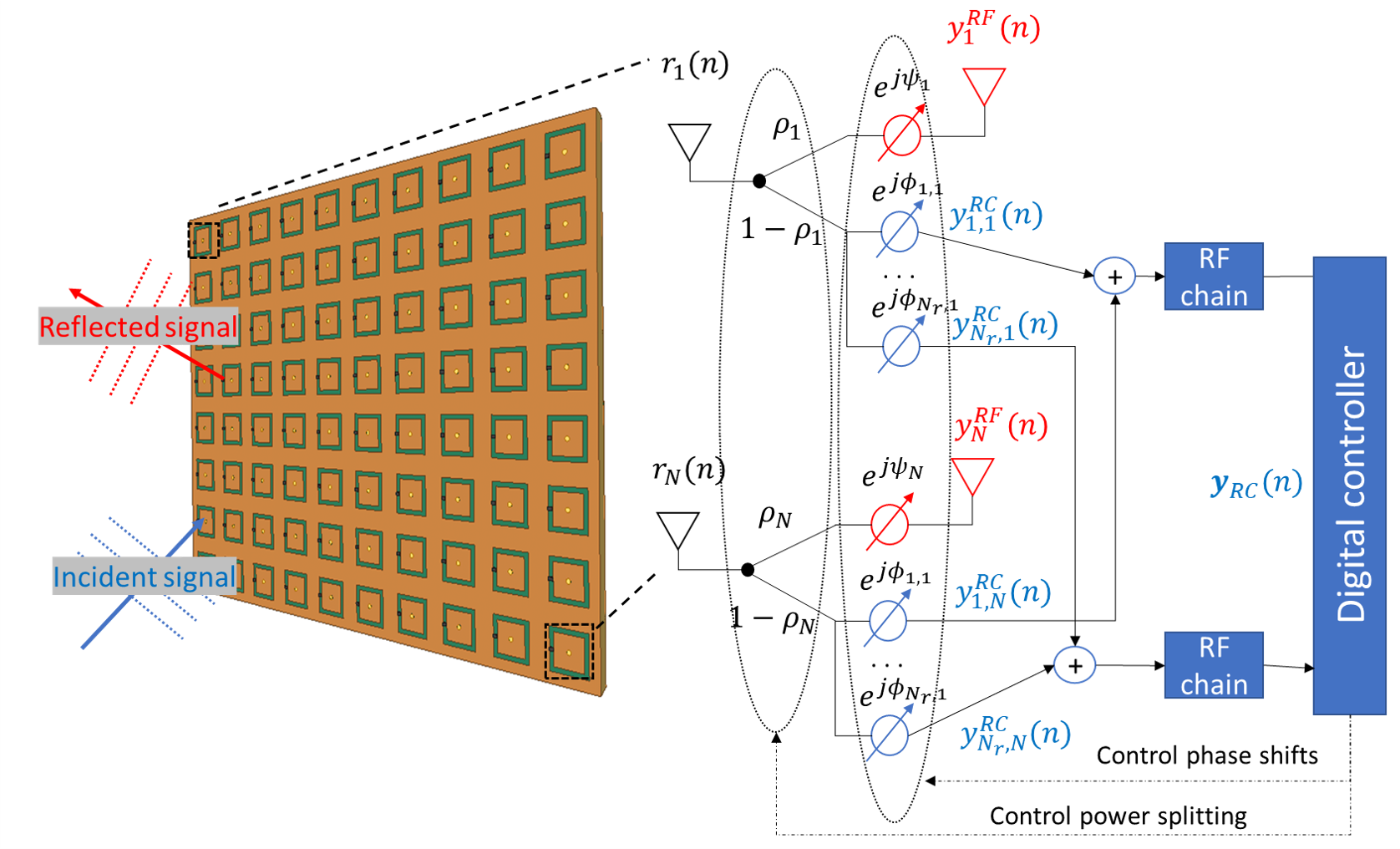}
		\vspace{-0.4cm}
		\caption{A simple model for the proposed \ac{hris} operation.} 
		\label{fig:Hybrid_RIS_model}
	\end{figure}

\vspace{-0.1cm}
To model the dual reflection-reception operation of HRISs, we consider a hybrid metasurface comprised $N$ meta-atom elements, which are connected to a digital controller via $N_r$ receive RF chains. Let $r_{l}(n)$ denote the radiation observed by the $l$-th HRIS element ($l=1,2,\ldots,N$) at the $n$-th time instance. A portion of this signal, dictated by the parameter $\rho_l \in [0,1]$, is reflected with a controllable phase shift $\psi_l \in [0,2\pi)$, and thus the reflected signal from the $l$-th element can be mathematically expressed as:
\begin{equation}\label{eq:reflection}
y_l^{\rm RF} (n)= \rho_l\,e^{\jmath \psi_l} r_l(n). 
\end{equation} 
The remainder of the observed signal is locally processed, via analog combining and digital processing. The signal forwarded to the $r$-th RF chain via combining, with $r \in \{1,2,\ldots,N_r\}$, from the $l$-th element is consequently given by 
\begin{equation}\label{eq:power_split_t}
y_{r,l}^{\rm RC} (n) = (1 - \rho_{l})e^{\jmath \phi_{r,l}} r_l(n), 
\end{equation}
where $\phi_{r,l}\in[0,2\pi)$ is the adjustable phase that models the joint effect of the response of the $l$-th meta-atom and the subsequent analog phase shifting. The proposed HRIS operation model is illustrated in Fig.~\ref{fig:Hybrid_RIS_model}. It is noted that the operation of the conventional passive and reflective RISs can be treated as a special case of our proposed \ac{hris}   architecture, by setting all $\rho_{l}$'s in \eqref{eq:reflection} equal to 1. Meanwhile, compared to existing relay techniques, there are two major advantages with our novel architecture. First, HRISs allow full-duplex operation (i.e., simultaneous reflection and reception) without inducing self interference, which is unavoidable in full-duplex relay systems. Second, HRISs require low power consumption since they do not need active power amplifiers.
	
The resulting signal model at HRIS can be expressed in vector form, as follows. By stacking the received $r_l(n)$ $\forall$$N$ and the reflected signals $y_l^{\rm RF} (n)$ $\forall$$N$ at the $N\times 1$ vectors $\myVec{r}(n)$ and  $\myVec{y}_{\rm RF}(n)$, respectively, it follows from \eqref{eq:reflection} that 
\begin{equation} \label{eq:reflection_vector}
    \myVec{y}_{\rm RF}(n) = { \boldsymbol  \Psi} \left( \boldsymbol \rho, \boldsymbol \psi\right) \myVec{r}(n),
\end{equation}
with $\myMat{\Psi} \left( \boldsymbol \rho, \boldsymbol \psi\right)  \triangleq {\rm diag} \left(\left[\rho_1\,e^{\jmath \psi_1},\rho_2\,e^{\jmath \psi_2},\dots,\rho_N\,e^{\jmath \psi_N}\right]\right)$.  
Similarly, by letting $\myVec{y}_{\rm RC}(n)\in \mathbb{C}^{N_r\times1}$ be the reception output vector at HRIS, it holds that
\begin{equation} \label{eq:output_compact}
    \myVec{y}_{\rm RC}(n) = {\boldsymbol \Phi}\left( \boldsymbol \rho, \boldsymbol \phi\right) \myVec{r}(n),
\end{equation}
where the $N_r \times N$ matrix $\myMat{\Phi}\left( \boldsymbol \rho, \boldsymbol \phi\right)$ represents the analog combining carried out at the \ac{hris} receiver. When the $l$-th meta-atom element is connected to the $r$-th RF chain, then $[\myMat{\Phi}]_{r,l} =  (1 - \rho_{l})e^{\jmath \phi_{r,l}}$, while when there is no such connection (e.g., for partially-connected combiners) it holds 
$[\myMat{\Phi}]_{r,l} = 0$. 

The reconfigurability of HRISs implies that the parameters $\rho_l$'s and the phase shifts $\psi_l$'s and $\phi_{r,l}$'s are externally controllable. It is noted that when an element is connected to multiple receive RF chains, then additional dedicated analog circuitry (e.g., conventional networks of phase shifters) is required to allow the signal to be forwarded with a different phase shift to each RF chain, at the possible cost of additional power consumption. Nonetheless, when each element feeds a single RF chain, then the model in Fig.~\ref{fig:Hybrid_RIS_model} can be realized without such circuitry by placing the elements on top of separated waveguides (see, e.g., \cite{shlezinger2020dynamic}).


\vspace{-0.2cm}
\subsection{HRIS-Assisted Channel Estimation}
\label{subsec:Problem}
\begin{figure}
    \centering
    \includegraphics[width=0.60\columnwidth]{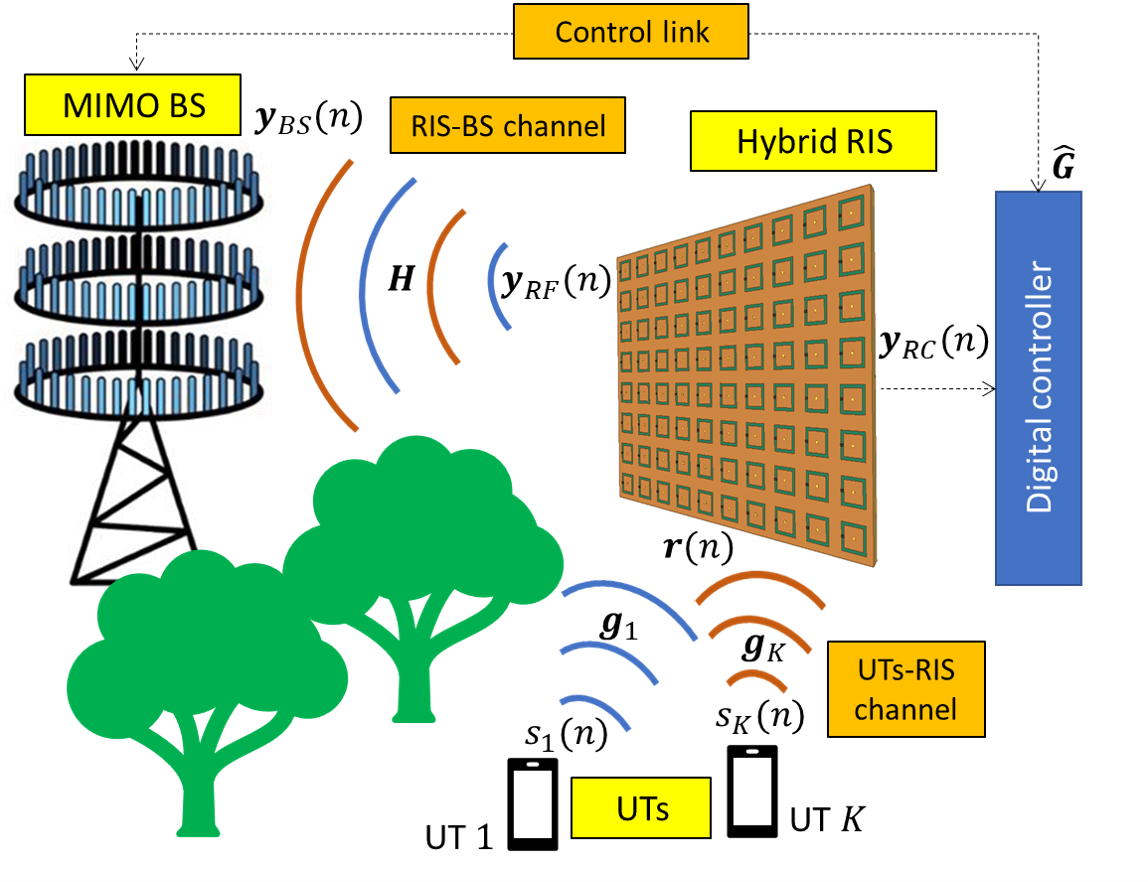}
    \caption{The considered \ac{hris}-empowered uplink multi-user \ac{mimo} system.}
    \label{fig:SystemModel1}
\end{figure}

\vspace{-0.1cm}
In order to investigate the capabilities of the proposed HRIS architecture in facilitating multi-user wireless communications, we henceforth consider the problem of channel estimation in \ac{hris}-empowered systems, being one of the main challenges associated with conventional nearly passive and reflective \acp{ris} \cite{hu2019two}. 
In particular, we consider an uplink multi-user \ac{mimo} system, where a \ac{bs} equipped with $M$ antenna elements serves $K$ single antenna \acp{ut} with the assistance of a  \ac{hris}. This setup is graphically presented in Fig.~\ref{fig:SystemModel1}.
We assume that there is no direct link between the \ac{bs} and any of $K$ \acp{ut}, and thus communication is done only via the \ac{hris}. Let  $\mathbf{H} \in \mathbb{C}^{M \times N}$ be the channel between the \ac{bs} and \ac{hris}, and ${\mathbf{g}}_k \in \mathbb{C}^{N}$ be the channel between the $k$-th user ($k=1,2,\ldots,K$) and \ac{hris}. We consider independent and identically distributed (i.i.d.) Rayleigh fading for all channels with $\mathbf{H}$ and ${\mathbf{g}}_k$ having i.i.d. zero-mean Gaussian entries with variances $\beta$ and $\gamma_k$, respectively, denoting the pathlosses.
 
Channel estimation with an HRIS can be carried out in a \acl{tdd} fashion using  $\tau$ orthogonal pilots $\{s_k(n)\}_{k=1}^K$ with $n=1,2,\ldots,\tau$. The signal observed by the \ac{hris} is given by the vector $\myVec{r}(n) = \mathbf{G} \myVec{s}(n)$ with $\mathbf{G}\triangleq \left[\mathbf{g}_1,\mathbf{g}_2,\dots,\mathbf{g}_K\right]$ and $\myVec{s}(n) \triangleq [s_1(n),s_2(n),\dots, s_K(n)]^T$. The signal locally processed by the \ac{hris} is given by \eqref{eq:output_compact} after being corrupted by \ac{awgn}, which models the thermal noise induced in signal acquisition. Consequently, the signals received at the \ac{hris} at the $n$-th channel estimation instance are given by:
 \begin{equation}\label{eq:received_all_pilot_output}
    \myVec{y}_{\rm RC}(n)={\boldsymbol \Phi}\left( {\boldsymbol \rho}(n), {\boldsymbol \phi}(n)\right) \mathbf{G} \myVec{s}(n) + \myVec{z}_r(n), 
\end{equation}
where $\myVec{z}_r(n) \in\mathbb{C}^{N_r}$ is the \ac{awgn} with entries of variance $\SigR$. Note that in \eqref{eq:received_all_pilot_output}, the configuration of the HRIS combining parameters, e.g.,  $\myVec{\rho}$ and $\myVec{\phi}$, are allowed to change over time. Similarly, the signal received at the \ac{bs} using \eqref{eq:reflection_vector} during the channel estimation phase can be expressed as:
\begin{equation}\label{eq:received_all_pilot_BS}
    \myVec{y}_{\rm BS}(n)= \mathbf{H} { \boldsymbol  \Psi} \left( {\boldsymbol \rho}(n), {\boldsymbol \psi}(n)\right) \, \mathbf{G} \myVec{s}(n) +  \myVec{z}_b(n),
\end{equation}
where $\myVec{z}_b(n) \in\mathbb{C}^{M }$ is \ac{awgn} with entries  of variance $\SigB$.
 
As in conventional \ac{ris}-empowered wireless networks, e.g., \cite{huang2019reconfigurable}, we assume that the \ac{bs} maintains a high-throughput direct link with the \ac{hris}. For passive \acp{ris}, this link is used for controlling the \ac{ris} reflection pattern. In \acp{hris} which have receiving capabilities, this link is also used for conveying information from the \ac{hris} to the \ac{bs}. Therefore, we focus on channel estimation carried out at both the \ac{hris} side as well as by the \ac{bs}. Our goal is to characterize the achievable \ac{mse} in recovering the \acp{ut}-\ac{hris} channel $\mathbf{G}$ from \eqref{eq:received_all_pilot_output}; along with the \ac{mse} in estimating $\mathbf{H}$ at the \ac{bs} from \eqref{eq:received_all_pilot_BS} and from the estimate of $\mathbf{G}$, denoted $\hat{\mathbf{G}}$, provided by the \ac{hris}.

\section{Identification of Individual Channels}
\label{sec:Estimation}

\vspace{-0.2cm}
\subsection{Channel Estimation without Noise}
\label{subsec:Noiseless}
\vspace{-0.1cm}
We begin by considering communications carried out in the ideal case, where the noise terms in \eqref{eq:received_all_pilot_output} and \eqref{eq:received_all_pilot_BS} are negligible, i.e., $\SigB,\SigR\rightarrow 0$. In such scenarios, one should be able to fully recover both $\mathbf{H}$ and $\mathbf{G}$ from the observed signals $\myVec{y}_{\rm RC}(n)$ and $\myVec{y}_{\rm BS}(n)$. The number of pilots required to achieve  accurate recovery is stated in the following proposition (its proof will be provided in the extended version of this paper):
%
\begin{proposition}
\label{pro:Noiseless}
In the high \ac{snr} regime, $\mathbf{H}$ and $\mathbf{G}$ can be accurately recovered when the number of pilots $\tau$ satisfies
\begin{equation}
    \label{eqn:Noiseless}
    \tau \geq N \cdot \max\left\{1,KN_r^{-1}\right\}.
\end{equation}
\end{proposition}

Proposition \ref{pro:Noiseless} reveals the intuitive benefit of \acp{hris} in facilitating channel estimation from reduced number of pilots, as compared to existing techniques for estimating the cascaded \acp{ut}-\ac{ris}-\ac{bs} channels (e.g., \cite{wang2020channel}). For instance, for a multi-user \ac{mimo} system with $M=16$ \ac{bs} antennas, $N_r=8$ \ac{hris} RF chains, $K=8$ \acp{ut}, and $N=64$ \ac{hris} elements, the adoption of an \ac{hris} allows recovering $\mathbf{H}$ and $\mathbf{G}$ separately using $\tau=64$ pilots. For comparison, the method proposed in \cite{wang2020channel} requires transmitting over $90$ pilots to identify the cascaded channel coefficients $\{[\mathbf{H}]_{m,l} [\mathbf{G}]_{l,k}\}, \forall, l,k$ and $m=1,2,\ldots,M$. This reduction in pilot signals is directly translated into improved spectral efficiency, as less pilots are to be transmitted in each coherence duration.

\vspace{-0.2cm}
\subsection{Channel Estimation with Noise}
\label{subsec:Noisy}
\vspace{-0.1cm}
The characterization of the number of required pilots in Proposition \ref{pro:Noiseless} provides an initial understanding of the \acp{hris}' capability in providing efficient channel estimation. However, as Proposition \ref{pro:Noiseless} considers an effectively noise-free setup, it is invariant of the fact that \acp{hris} split the power of the received signal $\myVec{r}(n)$ between the reflected and received components. In the presence of noise, this division of the signal power may result in \ac{snr} degradation. Therefore, we next study  channel estimation using \acp{hris} in the presence of noise. 

Let $\myVec{y}_{\rm RC}$ be the $\tau N_r\times 1$ vector generated by stacking $\myVec{y}_{\rm RC}(1),\myVec{y}_{\rm RC}(2),\ldots, \myVec{y}_{\rm RC}(\tau)$ and define $\myMat{\Phi}(n) \triangleq \myMat{\Phi}(\myVec{\rho}(n),\myVec{\phi}(n))$. It holds from \eqref{eq:received_all_pilot_output} that $\myVec{y}_{\rm RC}$ can be written as a linear function of the \acp{ut}-\ac{hris} channel $\mathbf{G}$, as follows:
\begin{equation}
    \label{eqn:LinearRC}
    \myVec{y}_{\rm RC} = \myMat{A}_{\rm  RC} {\rm vec}(\mathbf{G}) + \myVec{z}_r,
\end{equation}
where $\myVec{z}_r$ results from stacking $\myVec{z}_r(1),\myVec{z}_r(2),\ldots, \myVec{z}_r(\tau)$, while the entries of the matrix $\myMat{A}_{\rm  RC} \in \mathbb{C}^{\tau N_r \times KN}$ are given $\forall n,r,k$ and $i=1,2,\ldots,N$ by $\left[\myMat{A}_{\rm  RC}\right]_{(n-1)N_r + r, (k-1)N + i} \triangleq \left[\myMat{\Phi}(n)\right]_{r,i}s_k(n)$.
Similarly, by letting $\myVec{y}_{\rm BS}$ and $\myVec{z}_n$ be the stacking of $\myVec{y}_{\rm BS}(1),\myVec{y}_{\rm BS}(2),\ldots, \myVec{y}_{\rm BS}(\tau)$ and $\myVec{z}_{b}(1),\myVec{z}_{b}(2),\ldots, \myVec{z}_{b}(\tau)$, respectively, and defining $\myMat{\Psi}(n) \triangleq \myMat{\Psi}(\myVec{\rho}(n),\myVec{\psi}(n))$, we obtain
\begin{equation}
    \label{eqn:LinearBS}
    \myVec{y}_{\rm BS} = (\myMat{A}_{\rm BS} \otimes \myMat{I}_{M}) {\rm vec}(\mathbf{H}) + \myVec{z}_b.
\end{equation}
In \eqref{eqn:LinearBS}, the matrix  $\myMat{A}_{\rm BS} \in \mathbb{C}^{\tau \times N}$ is consisting of the entries: $  [\myMat{A}_{\rm BS}]_{n,  i} \triangleq [\myMat{\Psi}(n)]_{i,i}\sum_{k=1}^K [\mathbf{g}_k]_i s_k(n)$.

We next use the latter expressions formulations to quantify the achievable \ac{mse} in estimating the individual \acp{ut}-\ac{hris} and \ac{hris}-\ac{bs} channels for a fixed \ac{hris} configuration $\myVec{\rho}$, $\myVec{\phi}$, and $\myVec{\psi}$. We focus here on the case where these parameters remain constant during the channel estimation phase, and the noise powers at the \ac{hris} and \ac{bs} are of the same level, i.e., $\SigB = \SigR=\sigma^2$.
We begin by characterizing the achievable \ac{mse} performance in recovering $\mathbf{G}$, denoted by $\mathcal{ E}_{\rm G} \left( \left\{\boldsymbol \rho(n), \boldsymbol \phi(n) \right\}\right) $.
\begin{theorem}\label{Theorem:1}
The \acp{ut}-\ac{hris} channel $\mathbf{G}$ can be recovered with the following \ac{mse} performance:
\begin{equation*}
\mathcal{ E}_{\rm G} \left( \left\{\boldsymbol \rho(n), \boldsymbol \phi(n) \right\}\right)  = {\rm Tr} \left\{\left({\bf R}_g^{-1}\, + { \Gamma} \,\myMat{A}_{\rm  RC}^H\, \myMat{A}_{\rm  RC}\right)^{-1}\right\},
\end{equation*}
where ${\bf R}_g \triangleq {\rm blkdiag}\left\{\gamma_1{\bf I}_{N},\gamma_2{\bf I}_{N},\ldots ,\gamma_K{\bf I}_{N}\right\}$ and $\Gamma\triangleq\frac{P_t}{\sigma^2}$, with $P_t$ denoting each UT's transmit power for pilots.
\end{theorem}

Theorem~\ref{Theorem:1} allows to compute the achievable \ac{mse} in estimating $\mathbf{G}$ at the \ac{hris} side for a given configuration of its reception phase profile, determined by $\boldsymbol \rho \left(n \right)$ and $\boldsymbol \phi \left(n \right)$. The estimated $\mathbf{G}$ will be conveyed to the \ac{bs} (via their control link), and is used along with the observed reflected by the \ac{hris} to recover $\mathbf{H}$. 
When $\mathbf{G}$ is accurately estimated, the \ac{bs} can recover $\mathbf{H}$ up to the \ac{mse} stated in  the following theorem. 

\begin{theorem}\label{Theorem:2}
The \ac{hris}-\ac{bs} channel $\mathbf{H}$ can be recovered with the following \ac{mse} performance with ${\bf R}_h\triangleq\beta \myMat{I}_{M \tau}$: 
\begin{align*}  
&\mathcal{E}_H\left( \left\{\boldsymbol \rho \left(n \right),  \boldsymbol \psi \left(n \right), \boldsymbol \phi \left(n \right)\right\} \right) \\
&= {\rm Tr} \bigg(\Big( {\bf R}_h^{-1} + \Gamma  (\myMat{A}_{\rm BS} \otimes \myMat{I}_{M})^H(\myMat{A}_{\rm BS} \otimes \myMat{I}_{M})\Big)^{-1}\bigg).
\end{align*}
\end{theorem}

Theorems~\ref{Theorem:1} and \ref{Theorem:2}, whose proofs will be given in the extended version of this paper, allow to  evaluate the achievable \ac{mse} in recovering the individual channels $\mathbf{G}$ and $\mathbf{H}$. The fact that these \acp{mse} are given as functions of the \ac{hris} parameters $ \left\{\boldsymbol \rho \left(n \right),  \boldsymbol \psi \left(n \right), \boldsymbol \phi \left(n \right)\right\}$ enables us to numerically optimize its configuration. 
In Section~\ref{sec:Sims}, our numerical evaluation of the \acp{mse} reveals the fundamental tradeoff between the ability to recover $\mathbf{G}$ and $\mathbf{H}$, which is dictated mostly by the parameter $\mathbf{\rho}$ determining what portion of the impinging signal is reflected. 


\vspace{-0.2cm}
\subsection{Discussion}
\label{subsec:Discussion}
\vspace{-0.1cm}
The fact that \acp{hris} require less pilots naturally follows from their ability to provide additional $N_r$ receive ports, while simultaneously acting as a reflector.  It is noted that our results in the previous subsections are obtained assuming that the \acp{ut}-\ac{hris} channel $\mathbf{G}$ is estimated at the \ac{hris}, and its estimate is forwarded to the \ac{bs}. For this reason, Proposition~\ref{pro:Noiseless}  relies on having the \ac{hris} configuration change over time, as for static $\myVec{\rho}$ and $\myVec{\phi}$, one can only recover $\myMat{\Phi}\mathbf{G}$ via \eqref{eq:received_all_pilot_output}, from which $\myMat{G}$ cannot be computed when the \ac{hris} has less RF chains than elements, i.e., $N_r < N$. 
Furthermore, exploiting the  \ac{hris} as an additional non-co-located receive port can also facilitate data transmission once the channels are estimated, though in this case one would also have to account for possible rate limitations on the \ac{hris}-\ac{bs} link. We leave the study of these additional usages of \acp{hris} for future research. 

The majority of the existing literature about channel estimation in \ac{ris}-empowered communications has mainly focused on estimating the cascaded channel, which represents the joint effect of the \ac{ut}s-\ac{ris} and \ac{ris}-\ac{bs} channels in an entangled manner. Estimating the cascaded channel using reflective \acp{ris} typically requires a large amount of pilots in each coherence duration of the channel. Furthermore, knowledge of the individual channels enables the design of flexible and improved transmission and management schemes, compared to knowing solely the cascaded channel \cite{hu2019two}. In fact, several hardware architectures were proposed to provide some information processing to be carried out at the \ac{ris} side, by adding dedicated reception-only hardware \cite{taha2021enabling}-\cite{alexandropoulos2020hardware},
The  \ac{hris} architecture balances reception and reflection  in a controllable manner. In particular, \acp{hris} can  be configured to be divided into purely reflective and purely receptive elements, as in \cite{alexandropoulos2020hardware}, while providing additional degrees of freedom due to the ability to adjust the power splitting coefficients $\myVec{\rho}$.  

The study of \acp{hris}, combined with the numerical evaluations in Section~\ref{sec:Sims}, only reveal a portion of the potential of  \acp{hris} in facilitating wireless communication over programmable environments.  To further understand the contribution of \acp{hris}, one should also study their impact on data transmission, as well as consider the presence of an additional direct channel between the \acp{ut} and the \ac{bs}. Furthermore, the simplified model used in this work is based on the hybrid metamaterial model 
presented in \cite{alexandropoulos2021hybrid}, and additional experimental studies of this model are required to formulate a more accurate physically-compliant model for the behavior of \acp{hris}. These  extensions are left for future work.


\vspace{-0.2cm}
\section{Numerical Results}
\label{sec:Sims}
\vspace{-0.1cm}

In this section, we numerically evaluate the channel estimation performance of the proposed \ac{hris}-empowered multi-user MIMO systems. In our simulations, a \ac{bs} with $M=16$ antennas serves $K=8$ \acp{ut} via an \ac{hris} with $N=64$ elements.  The pathlosses of the individual channels ${\bf H}$ and ${\bf g}_k$ are modeled as $\beta = \lambda_0 \left(\frac{d_H}{d_0}\right)^{-\alpha_h}$ and $\gamma_k = \lambda_0\left(\frac{d_k}{d_0}\right)^{-\alpha_g}$, where $\lambda_0 = -20$ dB denotes a constant pathloss at the reference distance $d_0 = 1$ m, while $d_H$ and $d_k$ are the distances from the \ac{hris} to \ac{bs} and the $k$th \ac{ut}, respectively. The pathloss factors were set as  $\alpha_h = 2.2$  and $\alpha_g = 2.1$.  We consider a 2D Cartesian coordinate system in which the BS and the \ac{hris} are respectively located at points (0, 0) and (0, 50 m), while the $K$ users are randomly generated in an area centered at (30 m, 50 m) with a radius of $10$ m. 

\begin{figure}
    \centering
    \includegraphics[width=0.62\columnwidth]{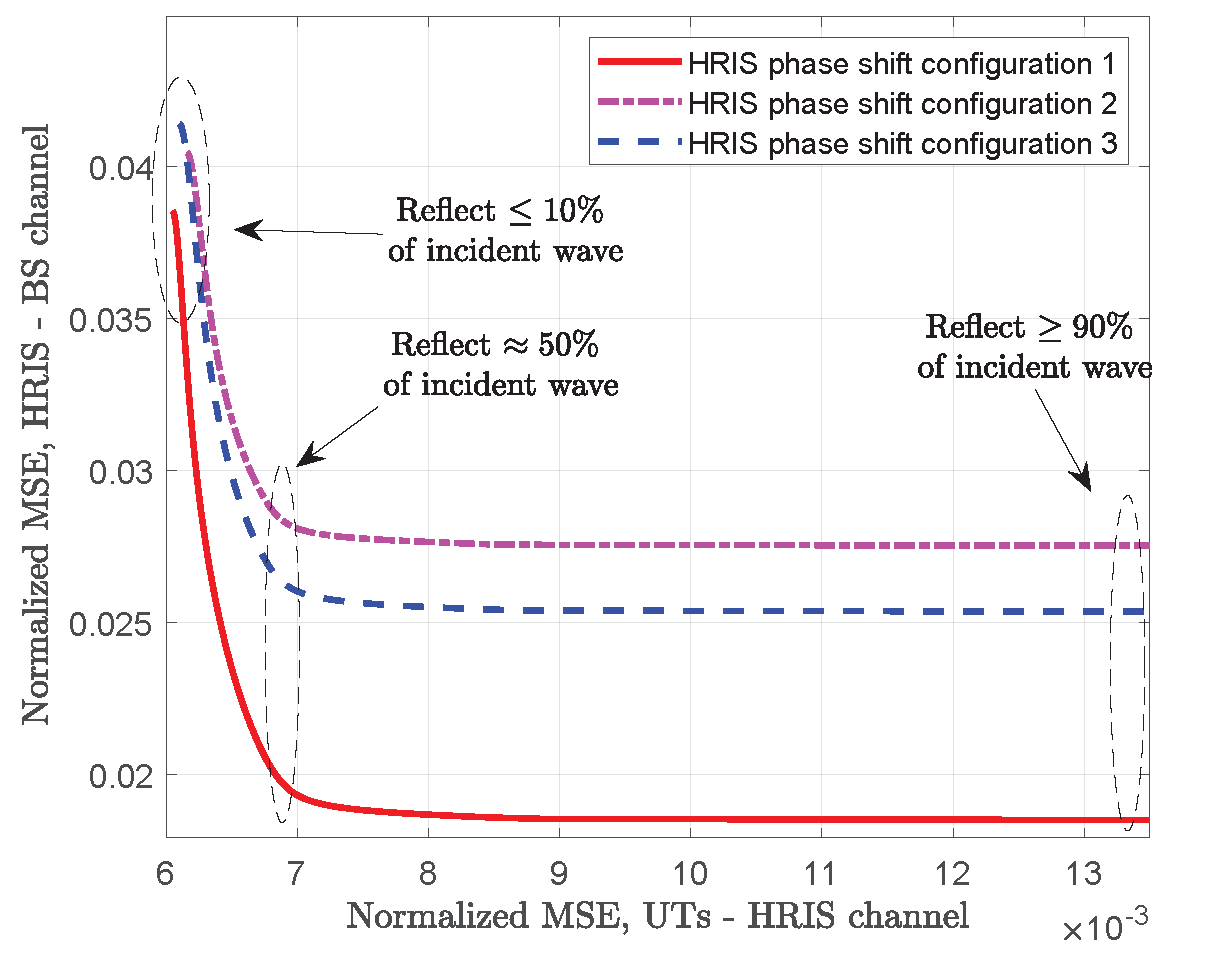}
    \caption{Normalized \ac{mse} performance in recovering the combined \acp{ut}-HRIS channel at the HRIS and the HRIS-\ac{bs} channel at the \ac{bs} with $\tau=70$.
    }
    \label{fig:tradeoff}
\end{figure}

In Fig.~\ref{fig:tradeoff}, we show the trade-off between the normalized \ac{mse} performances when estimating the \acp{ut}-\ac{hris} channel at the \ac{hris} and the \ac{hris}-\ac{bs} channel at the \ac{bs} for different values of the power splitting parameter $\rho$ and $30$ dB of transmit SNR. Each curve corresponds to a different random setting of the individual phase shift at each HRIS meta-atom element. As observed, there exists a clear trade-off between the accuracy in estimating each of the individual channels, which is dictated by how the \ac{hris} splits the power of the impinging signal. While the \ac{mse} values depend on the HRIS phase configuration, we observe that increasing the portion of the signal that is reflected in the range of up to $50\%$ notably improves the ability to estimate the \ac{hris}-\ac{bs} channel, while having only a minor effect on the \ac{mse} in estimating the \acp{ut}-\ac{hris} channel. However, further increasing the amount of power reflected, notably degrades the \ac{mse} in estimating the \acp{ut}-\ac{hris} channel, while hardly improving the accuracy of the \ac{hris}-\ac{bs} channel estimation.

\begin{figure}
    \centering
    \includegraphics[width=0.62\columnwidth]{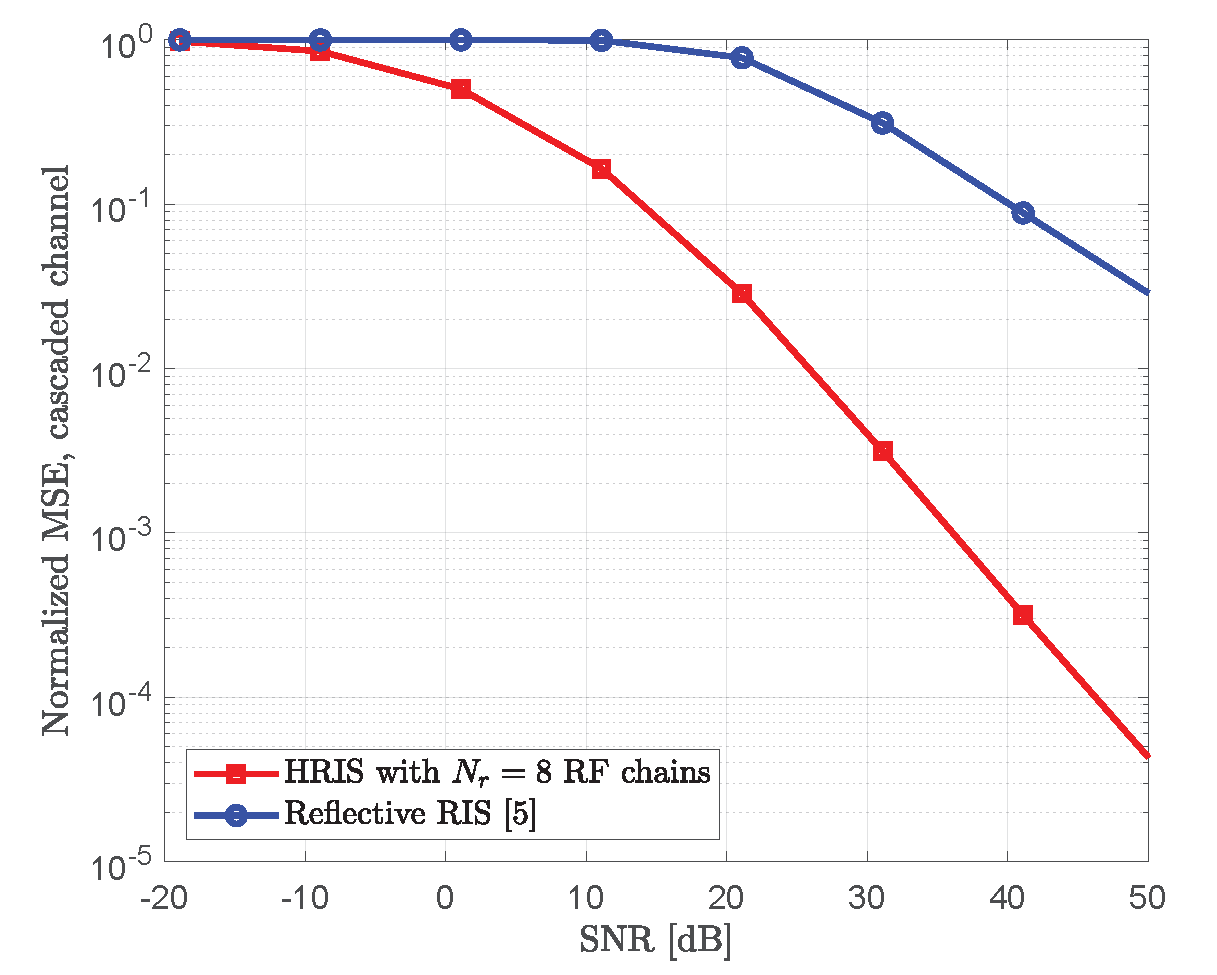}
    \caption{Normalized \ac{mse} performance in estimating the cascaded channel via an HRIS and a purely reflective RIS versus the transmit SNR in dB.
    }
    \label{fig:comparison}
\end{figure}

The normalized \ac{mse} performance in estimating the cascaded channel for the setup in Fig.~\ref{fig:tradeoff} with an \ac{hris}, which reflects on average {$50\%$} of the received signal, is compared to the method of \cite{wang2020channel} for reflective \acp{ris} in Fig. \ref{fig:comparison}. The cascased channel of our \ac{hris} is calculated based on the individually estimated \acp{ut}-\ac{hris} and \ac{hris}-\ac{bs} channels. The method in \cite{wang2020channel} requires over $90$ pilots, and thus we set the pilot length $\tau=100$. As shown in the figure, the \acp{hris} sensing capability is translated into improved cascaded channel estimation accuracy, compared to the state of the art. In particular, the considered \ac{hris} with $8$ receive RF chains achieves an \ac{snr} gain of over $20$ dB, though at the cost of higher power consumption and hardware complexity. 

\vspace{-0.2cm}
\section{Conclusions}
\label{sec:Conclusions}
\vspace{-0.1cm} 
In this paper, we presented the novel concept of HRISs, which are metasurfaces capable of simultaneously reflecting and sensing impinging signals in a dynamically controllable manner. We presented a simple model for their operation in HRIS-empowered multi-user MIMO communications systems, and investigated their potential to facilitate channel estimation, as an indicative application. We showed that at high SNRs, HRISs enable to significantly save pilot overhead compared to that required by purely reflective \acp{ris}. We also quantified the achievable estimation error performance for noisy channels. Our simulation results showcased the impactful role HRISs in RIS-empowered communications in estimating the individual channels as well as the cascaded channel over existing methods relying on nearly passive and reflective \acp{ris}.

\bibliographystyle{IEEEtran}
\bibliography{IEEEabrv,refs}

\end{document}